\documentclass[]{aa}

\usepackage{amsmath}
\usepackage{graphicx}
\usepackage{amssymb}
\usepackage{natbib}
\usepackage{bm}
\usepackage{longtable}
\usepackage{bbold}
\usepackage{url}
\usepackage{color}
\bibpunct{(}{)}{;}{a}{}{,}

\newcommand{\alphabold}{\mbox{\boldmath$\alpha$}}

\usepackage{graphicx}

\begin{document}

\title{A PCA approach to stellar effective temperatures 
\thanks{Tables 2, 4 and 5 are only available in electronic form
at the CDS via anonymous ftp to cdsarc.u-strasbg.fr (130.79.128.5)
or via http://cdsweb.u-strasbg.fr/cgi-bin/qcat?J/A+A/} 
}
\author{
Juli\'an Mu\~noz Bermejo \inst{1} \and 
Andr\'es Asensio Ramos \inst{1}\fnmsep\inst{2} \and 
Carlos Allende Prieto \inst{1}\fnmsep\inst{2}
}

\offprints{jbermejo@iac.es}

\institute{Instituto de Astrof\'{\i}sica de Canarias, 38205, La Laguna, Tenerife, Spain
\and
Departamento de Astrof\'{\i}sica, Universidad de La Laguna, 38205 La Laguna, Tenerife, Spain} 
\date{}

\titlerunning{A PCA approach to stellar effective temperatures}
\authorrunning{Mu\~noz Bermejo}
%
%
\abstract
{The derivation of the effective temperature of a star is a critical first step
in a detailed spectroscopic analysis. Spectroscopic methods
suffer from systematic errors related to model simplifications. Photometric methods
may be more robust, but are exposed to the distortions caused by interstellar reddening.
Direct methods are difficult to apply, since fundamental data of high accuracy are
hard to obtain.}
{We explore a new approach in which the spectrum is used to characterize a
star's effective temperature based on a calibration established by a small
set of standard stars.}
{We perform principal component analysis on homogeneous libraries of stellar spectra,
then calibrate a relationship between the principal components and the effective 
temperature using a set of stars with reliable effective temperatures.}
{We find that our procedure gives excellent consistency when spectra from
a homogenous set of observations are used. Systematic offsets may appear
when combining observations from different sources.
Using as reference the spectra of
stars with high-quality spectroscopic temperatures in the Elodie library, we
define a temperature scale for FG-type disk dwarfs with an internal 
consistency of about 50 K, in excellent agreement with 
temperatures from direct determinations  and widely used 
scales  based on the infrared flux method.}
{}

\keywords{Stars: fundamental parameters --- Catalogs --- Techniques: spectroscopic}

\maketitle
\email{jbermejo@iac.es}
\section{Introduction}
\label{Introduction}
The photons we measure from a star provide information on the regions they 
were last emitted or scattered from, i.e. the atmosphere of the star.
Accordingly, it is this shallow layer that matters for
modeling and understanding a stellar spectrum. 
Stellar spectra are described with three basic atmospheric
parameters: effective temperature ($T_{\rm eff}$), 
surface gravity ($\log g$), and overall metallicity ([Fe/H]). 
The effective temperature corresponds to the temperature of a 
black body with the same total radiative energy of the star:
\begin{equation}
F = \int_{-\infty}^{\infty} F_\nu d\nu = \sigma T_{\rm eff}^4.
\label{sb}
\end{equation}
\noindent In addition to the basic atmospheric parameters, we 
can infer a lot of information about a star from its spectrum,
most notably its chemical composition
from the strength of absorption lines associated with different
elements.

There are several ways to approach the task of determining  
the effective temperature of a star. One is by comparing
the observed spectral energy distribution with model 
calculations \citep[see, e.g.,][]{Ramirez}. 
Another possibility consists in using  
photometric calibrations, such as those based on the 
infrared flux method \citep{Blackwell,AlonsoDwarfs, Alonso1, Casagrande}. 

The flux at the stellar surface ($F$ at a distance
equal to the stellar radius $R$) and that at the Earth ($f$ at a distance
$d$ from the star) satisfy
\begin{equation}
F R^2 = f d^2.
\end{equation}
\noindent Based on this property, the most straightforward method
of deriving effective temperatures corresponds to measuring the 
bolometric flux of a star and its 
angular diameter ($2R/d$) and to using them with the Stefan-Boltzmann law 
(Eq. \ref{sb}). This method is, however, technologically challenging
due to the  tiny apparent sizes of stars in the sky -- the nearest
solar-like stars at merely a few pc from us are only
a few milliarcseconds in diameter. Nevertheless, the Center for 
High Angular Resolution Astronomy (CHARA) 
array measurements \citep{Chara1,Chara2} and other projects \citep{Mozurkewich,kervella} 
have provided us with angular diameters
with unprecedented quality for many giants and a handful
of dwarf stars.  

In this paper we introduce a new method of inferring the effective 
temperature of a star. 
Our procedure is inspired  by the work by  \citet{Cayrel}, who derive
effective temperatures by modeling the H$\alpha$ line, correcting for
systematic errors by linearly mapping their temperatures to a scale based on 
angular diameters and absolute fluxes. We condense the information on
 $T_{\rm eff}$ contained in stellar spectra using principal component 
 analysis (PCA). Then we map the principal components onto $T_{\rm eff}$s
based on a set of reliable calibration stars.
Once the calibration is performed, we can 
easily derive temperatures for other stars observed with
the same instrument. 

Since our procedure takes as input data continuum-normalized 
high-resolution spectra, it is immune to distortions in the spectral
energy distributions due to  interstellar reddening. 
Unlike $\chi^2$-fitting certain parts of the spectrum with 
high sensitivity to Teff, such as the Balmer lines, by using models, 
PCA is optimized to make use of all the information on the stellar 
effective temperature contained in the spectrum.

\begin{figure*}[t!]
\centering
\includegraphics[width=\textwidth]{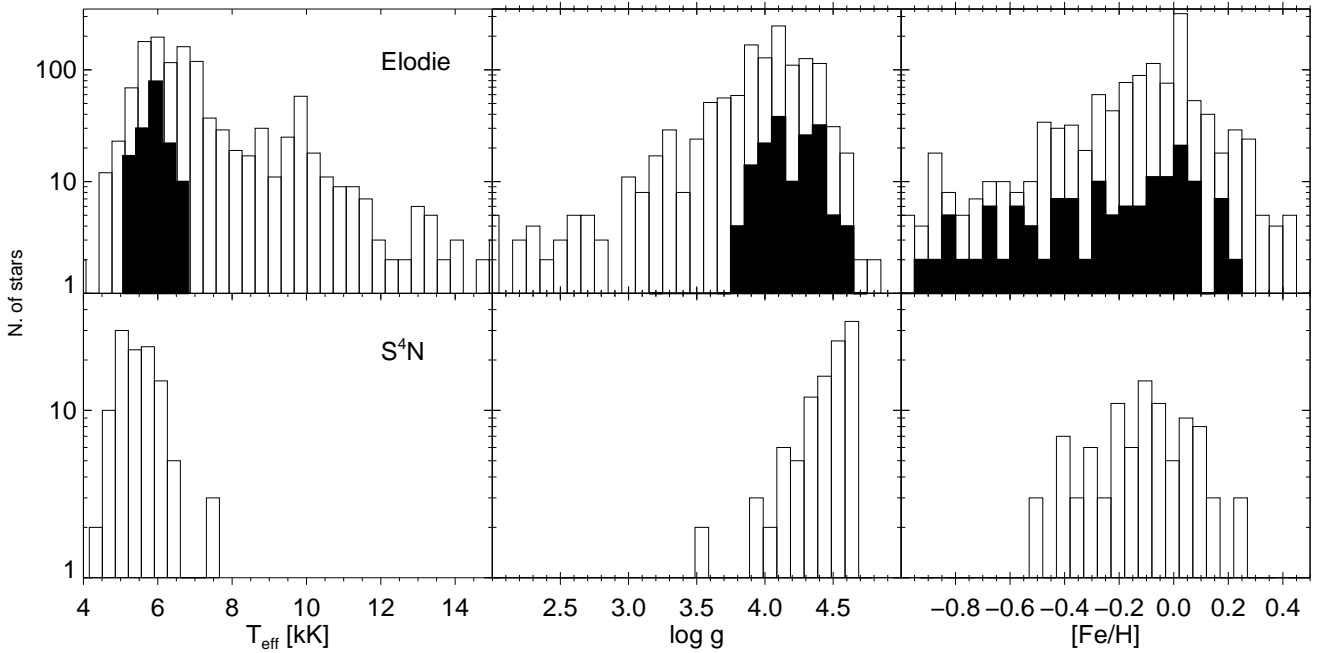}
\caption{Elodie (upper panel) and S$^4$N (lower panel) stellar parameters 
for all spectra in the sample. In the upper panels, the subsample of the Elodie
library with the highest quality effective temperatures ($Q_{T_\mathrm{eff}}=4$) 
is shown with filled-in black bars.}
\label{fig:survey_hist}
\end{figure*}

This paper is divided into five sections. 
Section \ref{The spectral libraries} provides details of the 
libraries that we use in our work, and the PCA analysis we performed
on them. Section \ref{sec:calibration} describes how we calibrate
a relationship between PCA coefficients and stellar 
effective temperatures.
Section \ref{Results} presents our 
results and Sect. \ref{Temperature Scales} compares them with other 
temperature scales from the literature. In Sect. \ref{allelodie}, we
apply our preferred PCA transformation to more than 18,000 spectra from
the Elodie archive. Finally, Sect. \ref{Conclusions} 
gives a summary of the work and our conclusions.

\section{The spectral libraries}
\label{The spectral libraries}
Among the many spectral libraries available in the literature, 
we have selected two that cover quite homogeneously the entire visible range 
(from 390 to 680 nm): Elodie and S$^4$N.
As described below, these two libraries differ fundamentally in two aspects: 
their spectra were obtained with 
different instruments and processed independently, 
and they focus on different types of stars. S$^4$N considers only
stars in the immediate solar neighborhood, with distances to the Sun smaller than 
about 15 pc, whereas Elodie includes more distant stars.

\subsection{Elodie}
\label{Elodie}
The Elodie library contains 1959 spectra of 1388 stars acquired with the Elodie
spectrograph installed in the 1.93 m telescope at the Observatoire de
Haute-Provence (France) \citep{ElodieB,ElodieA}\footnote{For more detailed
information check the webpage
\url{http://www.obs.u-bordeaux1.fr/m2a/soubiran/elodie_library.html}}. The spectra
on the library have a signal-to-noise ratio (SNR) between 100 and 150. They cover the 
range between 3100 K to 50000 K in effective temperature, 
$-0.25$ to 4.9 in $\log g$ (with $g$ in cm s$^{-2}$) and $-3$ 
to $+1$ dex in [Fe/H]\footnote{[X/H] $= \log$ (N(X)/N(H)) - $\log$ (N(X)/N(H))$_\odot$, where
N(X) represents the number density of nuclei of the element X.}. 
Histograms of these quantities are presented 
in the upper panel of Fig. \ref{fig:survey_hist}.

The spectra have a resolving power $R=\lambda/\delta\lambda=42000$, with the flux
normalized to a pseudo-continuum. We have performed our analysis both at a resolving power of 
$R= 10000$ and $1000$, finding slightly better results at lower resolution.
 Smoothing and resampling the spectra can only destroy information, but
it is also possible that it helps reducing the impact of high-frequency  
instrumental distortions in the data.
Working at low resolution offers additional advantages, since                   
the computational effort required is signicantly smaller and 
opens up the application of the method to lower-resolution instruments.

In order to degrade the spectral resolution, we smear the spectra using 
Gaussian convolution 
\footnote{With the {\tt gconv} IDL code available at \url{http://hebe.as.utexas.edu/stools/}}.
Additionally, we ensure that all spectra have a consistent continuum normalization. 
Each spectrum is fitted with an 
8-th order polynomial. Data points for which
the residual between the original spectrum and the fit are beyond 0.5 standard 
deviations below 
or 3 above the mean are discarded. This procedure is iterated ten times 
until the fit is
close to the upper envelope of the spectrum. This process is arguably not optimal
for locating the true stellar continuum, but it is consistently applied to all spectra.

Additional filtering is necessary, since some of 
the spectra show unwanted features, such as emission peaks or 
instrumental distortions, or correspond to outliers 
(for instance, when temperature is outside the range considered
in our analysis).
To filter the data we first discard all spectra with 
emission peaks greater than 1.2 times the continuum 
and absorption features deeper than 0.1 times the continuum flux. 
After that we apply PCA to the spectra that have passed 
the first filter, and keep only those for which the 
difference between the spectra reconstructed with the
first 5 principal components and the original spectrum is under 5\%. 
This ensures that we are picking up ``regular'' stars which share
properties with the rest of objects in the sample. That leaves us 
with a final set of 1245 spectra. In what follows we refer to this set of
spectra as ``Elodie''.

The atmospheric parameters of the Elodie stars (from the online
database) have been obtained from a compilation of high resolution spectroscopic 
analyses in the literature. When multiple values are available 
from different sources, the library catalog adopts a weighted average, 
giving preference to data with smaller errors. 
The goal is to end up with a homogeneous set of values for all parameters.
Quality flags (Q) are assigned to each parameter value according to 
the level of agreement across different sources. The better the
agreement, the larger the quality flag. 
$Q_{T\mathrm{eff}}$ ranges from $-1$ to $4$; 
in this scale, 1 corresponds to the lowest and 4 to the highest quality, 
being $-1$ and 0 special cases for internal 
determination of the effective temperature\footnote{Determined 
with the TGMET software by \cite{Elodie1}.}, and values derived from $B-V$ 
colors\footnote{Assuming the empirical color-temperature relation for a main 
sequence star and neglecting interstellar extinction, using to 
the Tycho2 I/259 catalog.}, respectively.  

\subsection{S$^4$N}
\label{S4N}

The S$^4$N library by \cite{Allende} includes spectra obtained with the 
Tull spectrograph \citep{Tull95} on the 2.7 m telescope at the 
McDonald Observatory of the University of Texas at Austin, 
and FEROS, the Fiber-fed Extended Range Optical Spectrograph \citep{feros}
attached (at the time) to the 1.52 m telescope at the 
European Southern Observatory (La Silla, Chile). It includes 
119 spectra of 119 stars (including the spectrum
of the blue sky as a proxy for the Sun), which have a SNR of 150-600, and a 
resolving power $R\simeq50000$. It covers the range of 
$-0.9$ to 0.5 in metallicity, 1.9 to 4.7 in $\log g$
and 4158 K to 7646 K in effective temperature, as illustrated in the
lower panel of Fig. \ref{fig:survey_hist}.

For consistency with the Elodie dataset, we have used the same wavelength 
grid and the same smoothing algorithm in S$^4$N, resulting in a final resolution of $R = 1000$. 
Likewise, exactly the same continuum normalization process has been applied to the 
S$^4$N spectra.

The original atmospheric parameters for the  S$^4$N library stars 
were obtained using several methods. 
The effective temperatures were calculated with the infrared flux method (IRFM)
calibrations described by \cite{AlonsoDwarfs}. 
We have recalculated the temperatures with the more recent 
$B-V$ and $b-y$ calibrations by \cite{Casagrande}, adopting the latter as we 
explain in Section 5. 
The $\log g$ values were obtained with stellar evolution models, 
from the effective temperatures (derived from the Alonso et al. calibration) 
and the parallaxes measured by the Hipparcos mission \citep{Hipparcos}, which are
accurately known for all the S$^4$N stars. 
The metallicity was then obtained by spectroscopic means, 
using $T_\mathrm{eff}$ and $\log g$ as known parameters.

\subsection{Combination}
\label{Connection}

Both Elodie and S$^4$N are spectroscopic libraries obtained with echelle spectrographs. 
Nevertheless, the instruments and the data processing are different in each case.
 We first attempted to use the full spectral region discussed in the preceding
sections, but the results were significantly poorer than those for a single
library. Therefore, we found it important to choose appropriate spectral windows 
instead  of the entire wavelength range, so that the existing differences between 
libraries do not affect our results. 
The presence of the sought-after information in the spectra is a must,
but nearly every feature in the spectrum responds to {\rm Teff}, and
PCA is designed for the very purpose of condensing such information. The
discrepancies between the two libraries are dominated by 
instrumental or atmospheric (telluric) distortions.

The root mean squared error (rms) between two different stars in the same library
(all wavelengths) is about 3\%, 
whereas the mean difference between two different spectra of 
the same star in the same library is about 1\% (derived from stars observed more than
once in Elodie). 

\begin{figure}
\centering
\includegraphics[width=\columnwidth]{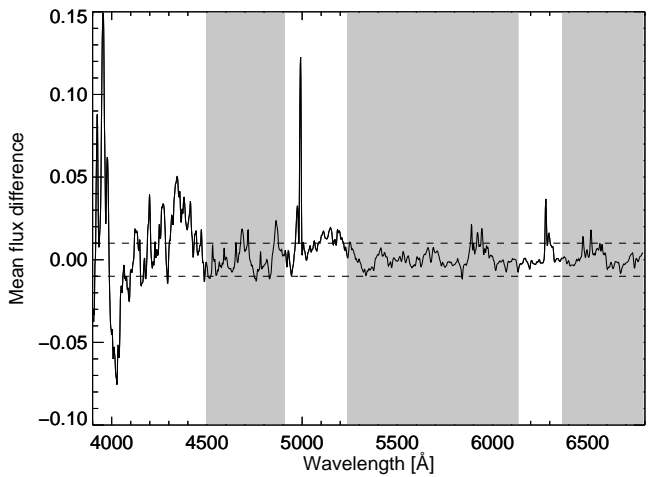}
\caption{Mean difference spectrum between Elodie and S$^4$N for the 29 stars in
common in both surveys. The dashed horizontal lines mark $\pm$1\% and the shadowed regions
indicate the regions that we consider for the combined analysis of both databases.}
\label{fig:range}
\end{figure}

To select the wavelength regions that we consider in common for both surveys, 
we apply the following
simple algorithm based on the fact that Elodie and S$^4$N have 29 stars (and the
same number of spectra) in common. Calculating the 
mean difference for those 29 spectra, we get what we can call a ``mean difference spectrum''
which is displayed in Figure \ref{fig:range}. 
Then, we selected the regions where this mean difference is small.
The mean rms of the spectra we have in common in our 
spectral range is 0.8 \% (under 1\%, which can be considered as a suitable threshold because this is
the largest difference between different spectra of the same star in one of the databases).
This criterion leads to the shaded regions in Fig. \ref{fig:range}, that correspond to the windows
[4449,4907] \AA, [5237,6134] \AA\ and [6368,6790] \AA.

\begin{figure*}
\centering
\includegraphics[width=\textwidth]{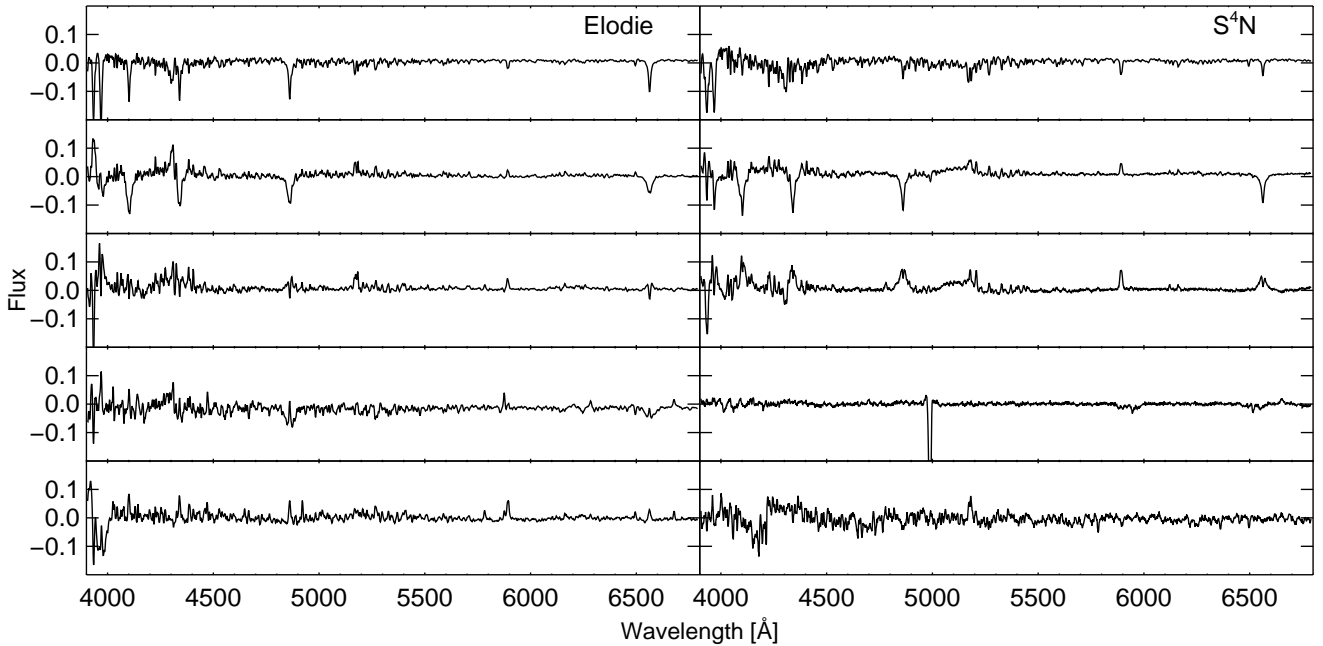}
\caption{First 5 eigenvectors (ordered top to bottom) 
of the Elodie (left column) and S$^4$N (right column) databases.}
\label{fig:Elodie_S4N_Evecs}
\end{figure*}

\begin{figure}
\includegraphics[width=0.88\columnwidth]{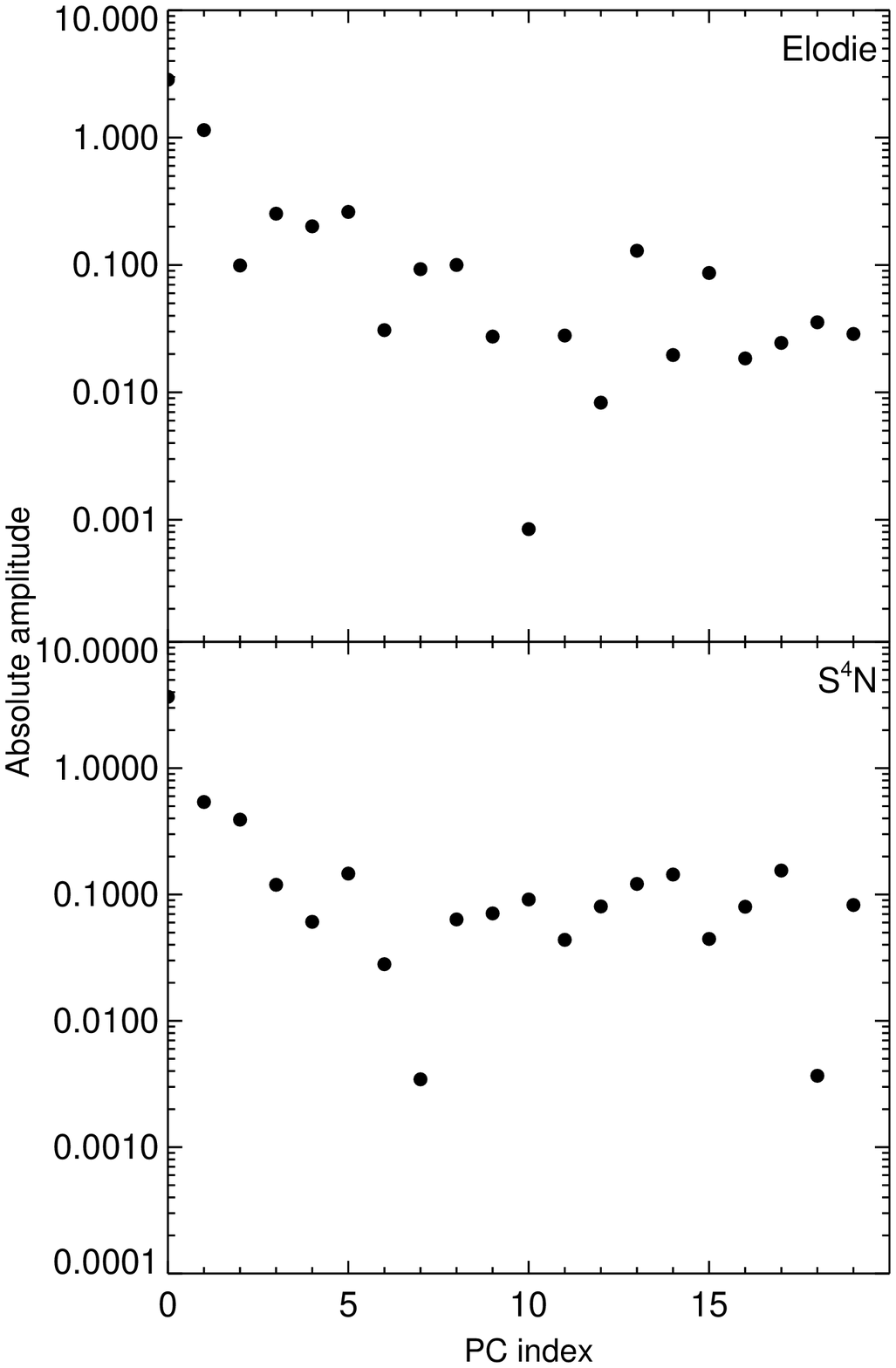}
\caption{First 20 principal components of the first star in each library: HD 245 for Elodie (top) 
and the Sun for S$^4$N (bottom).}
\label{fig:PCs}
\end{figure}

\subsection{PCA analysis}
\label{PCA application}

Principal component analysis is an algebraic and statistical tool which aims 
at finding  the directions of largest variance in the data.
Given a set of stellar spectra, we can apply PCA to them 
and describe each spectrum with far fewer numbers than the original data.  
In order to arrive at the new set of numbers (the principal components; PCs), 
we adopt a new basis set formed by the eigenvectors of the 
correlation matrix, and order them by decreasing eigenvalues. 
With just the  projection of the data into the first 10 elements of 
the base we can reproduce optical low resolution spectra with a very 
small error (less than 2\% in most cases).
Not only is PCA a powerful compressing algorithm, 
but it also gives us information about the data. 


We apply PCA and retain the first $N$ principal components 
for each star in the sample; we typically work with $\sim$10. 
Then we look for a calibration
between the principal components of a subset of stars 
(the calibrators) and their effective temperatures. Finally, 
we use that calibration to 
infer the temperature of the rest of stars (test).

\begin{figure*}
\centering
\includegraphics[width=\columnwidth]{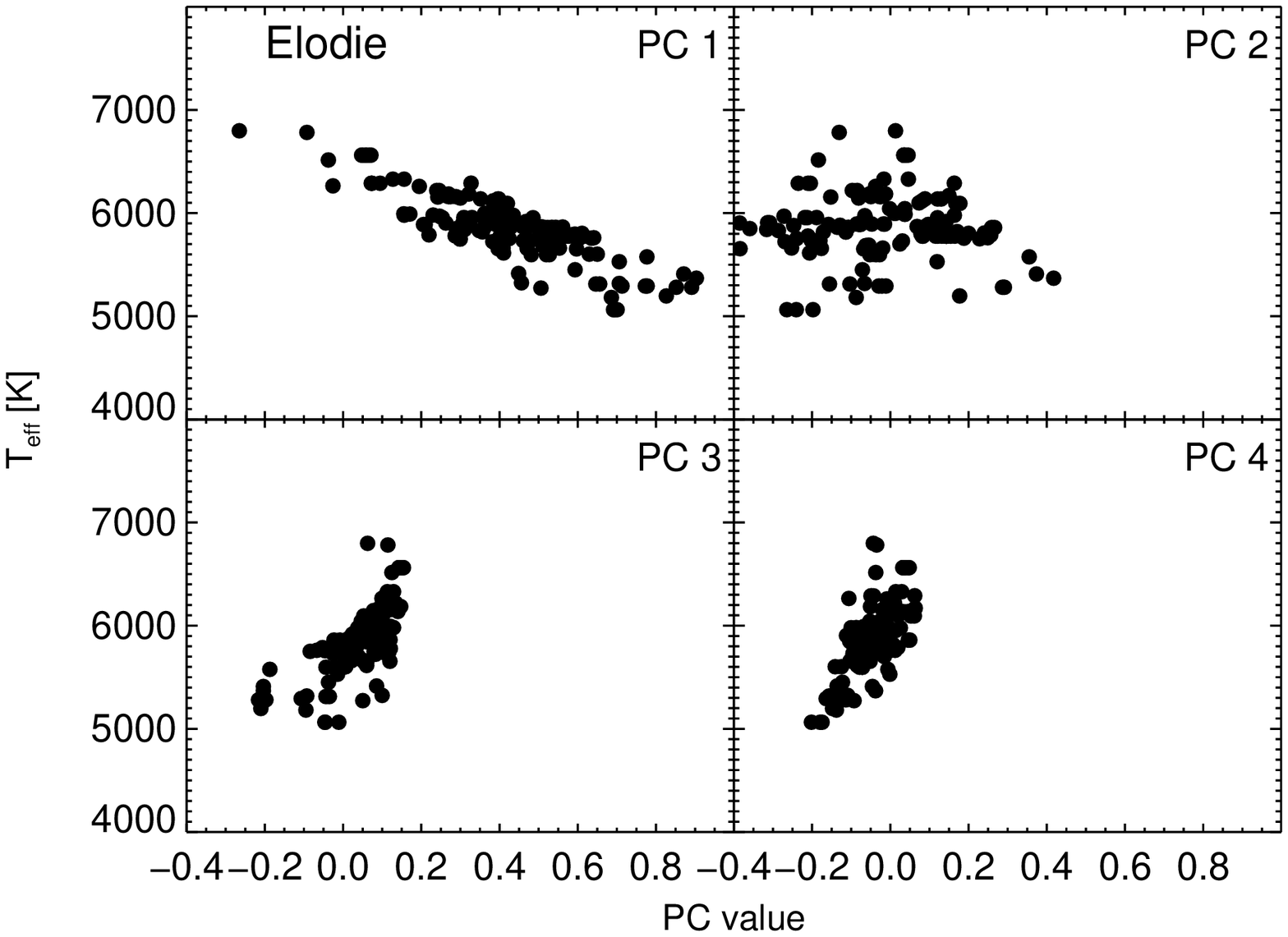}
\includegraphics[width=\columnwidth]{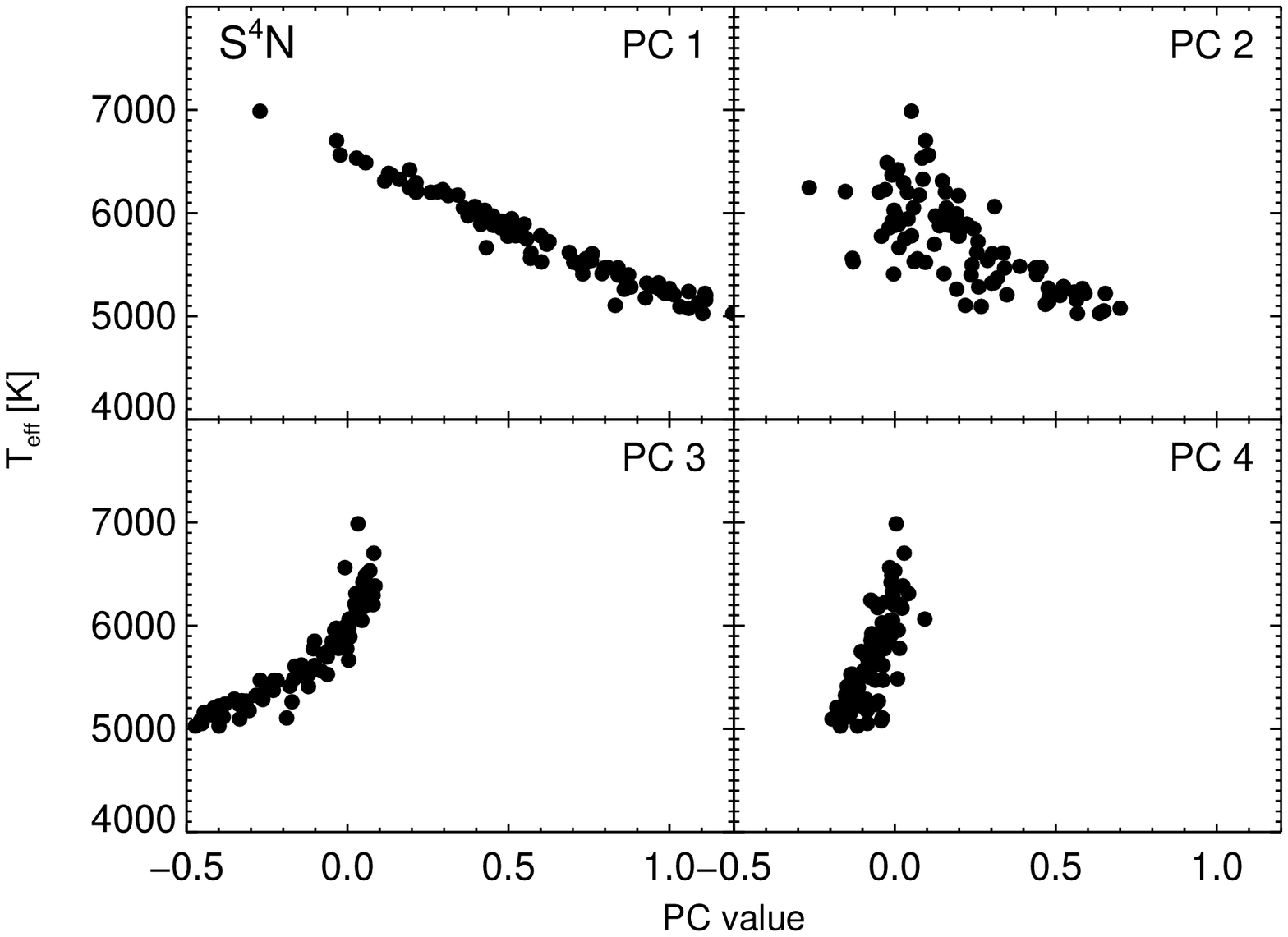}
\caption{Plot of the T$_\mathrm{eff}$ versus the first 4 PCs of every star of Elodie (left panels)
and S$^4$N (right panel).}
\label{fig:Teff_vs_PCs}
\end{figure*}

As a brief reminder of the procedure to compute the principal components, 
let us assume that the $m \times n$ matrix of data $\mathbf{Y}$ is built by stacking
as rows the spectra of size $m$ of all the $n$ stars considered, where the 
average spectrum
\begin{equation}
m(\lambda_j) = \frac{1}{n} \sum_{i=1}^n Y_i(\lambda_j),
\end{equation}
with $j=1\ldots m$, has been subtracted from each observation. From the zero-mean
data, we compute the correlation matrix:
\begin{equation}
\mathbf{C}=\mathbf{Y} \mathbf{Y}^T
\end{equation}
and we diagonalize it. Note that it may be desirable to diagonalize the
matrix $\mathbf{Y}^T \mathbf{Y}$ 
(along the observation direction, instead of the wavelength direction)
if this has smaller dimensions. It is simple to transform
back and forth from the eigenvectors in one representation to the other  
by appropriately multiplying by the data matrix $\mathbf{Y}$
 (see, e.g., the discussion in Mart\'{\i}nez Gonz\'alez et al. 2008).
The eigenvectors computed so far represent the directions in the 
space of spectra where we find the largest correlation. The first 5
eigenvectors obtained for Elodie and S$^4$N are displayed in Fig. \ref{fig:Elodie_S4N_Evecs}. 
Interestingly, we find that the fourth eigenvector of S$^4$N contains a 
conspicuous peak at $\sim 5000$ \AA. We believe this to be residuals from
narcissus in one of the spectrographs,  the {\it picket 
fence} described by \cite{Tull95}.

Figure \ref{fig:PCs} shows the first 20 principal components for the first star
of each library (HD 245 in the case of Elodie and the Sun in the case of S$^4$N). As usual, they tend to decrease because the first PCs
contain the most important information of the spectrum.
The calibration we develop below builds on 
the data shown in Fig. \ref{fig:Teff_vs_PCs}, which shows the relationships between
$T_\mathrm{eff}$ and the first five principal components. Since these relations 
are, in general, not  linear,
more complicated expressions are necessary. This is discussed in
Section Sect. \ref{sec:calibration}.


\section{Calibration}
\label{sec:calibration}

\subsection{General considerations}
Our goal is to find a suitable function that, given the principal components of a 
collection of stars, allows us to compute their effective temperature. 
Once that function has been established and tested for a calibration
set, we can apply it to stars with unknown effective temperatures.
There are three potential problems:
\begin{enumerate}
\item First, it is important to include stars with spectra of sufficient
quality. We verified that a calibration of the effective temperature using
the entire
Elodie sample (including stars with $Q_{T_\mathrm{eff}}<4$) leads to a poor calibration. 
We also believe that spectra with different quality flags are subject to 
different systematic offsets. To solve this issue, we only used 
stars with $Q_{T_\mathrm{eff}} = 4$, ending up with a fairly homogeneous 
temperature scale.

\item Second, we found difficulties in the simultaneous
calibration of stars over a broad range of temperatures. 
It is obvious that it is of paramount importance to have a sample of stars
spanning a sufficiently broad range of physical conditions, since otherwise 
the results will not be useful. Our calibration considers stars with 
temperatures between 5000 K and 7000 K and $\log g \geq 3$. 
In fact, there were only a few stars with temperatures outside this range in the
libraries considered. The $\log g$ range was set to avoid giants; 
given their scarcity
in our sample, they would cause a degradation in the calibration.
After imposing the selection criteria described above, we ended up with 
159 spectra in Elodie and 86 in S$^4$N.

\item Third, we found the application of standard regression 
algorithms inadequate. The final calibration procedure has to
be general enough to be applied to stars of unknown effective temperature.
Given that we
include a large number of PCs in the regression, a regular linear 
regression algorithm based on a maximum-likelihood approach
results in overfitting; it not only follows generic properties of
the stars  as representative examples of their
classes, but it also includes peculiarities of individual stars
in the calibration sample. 
To address this issue we employed a Bayesian non-parametric 
regression algorithm based on a Relevance Vector Machine \cite[RVM;][]{Tipping}, 
which uses Bayesian inference to learn about the data.
\end{enumerate}

\subsection{Bayesian calibration}
\label{Actual Calibration}

Given that the RVM avoids overfitting, we propose a sufficiently
general functional form for the calibration, and let the data decide on
is the optimal level of 
complexity for our sample. To this end, we write the
effective temperature as 
\begin{equation}
T_\mathrm{eff} =T_0 + \sum_{j=1}^C a_j \mathrm{PC}_j + \sum_{j=1}^C b_j \mathrm{PC}_j^2 + \sum_{j=1}^C c_j \mathrm{PC}_j^3 + \sum_{j=1}^C d_j \mathrm{PC}_j^4,
\label{eq:calibration}
\end{equation}
where PC$_j$ is the $j$-th principal component of a given star, $C$ is the number of principal
components that we consider, and $T_0$, $a_j$, $b_j$, $c_j$ and $d_j$ are coefficients that we have to
infer from the calibration data. For the sake of simplicity, we use a compact notation in which
the vector $\mathbf{w}$ of length $4C+1$ is built by stacking all the coefficients together
\begin{equation}
 \mathbf{w} = (T_0,a_1,a_2,...,a_C,b_1,...,d_C).
\end{equation}

The RVM is based on a Bayesian hierarchical approach to linear regression. The aim is to use the 
available data to compute the posterior distribution function 
for the vector of weights $\mathbf{w}$ and the noise variance $\sigma^2$ (that can even be
estimated from the same data). Therefore, a direct application of the Bayes theorem yields the
posterior distribution function for the unknowns
\begin{equation}
p(\mathbf{w},\sigma^2|\mathbf{d}) = \frac{p(\mathbf{d}|\mathbf{w},\sigma^2) p(\mathbf{w},\sigma^2)}{p(\mathbf{d})},
\end{equation}
where $\mathbf{d}$ is the data, which contains the principal components and the
effective temperature, $p(\mathbf{d}|\mathbf{w},\sigma^2)$ is the likelihood function 
that gives an
idea of how well the model fits the data, $p(\mathbf{w},\sigma^2)$ is the
prior distribution for the parameters and 
$p(\mathbf{d})$ is the evidence \citep[e.g.,][]{Gregory}. 
The key ingredient invoked by \cite{Tipping} is to build a hierarchical prior for $\mathbf{w}$. 
The prior for  $\mathbf{w}$ will depend on a set of hyperparameters $\alphabold$, 
which are learned from the data during the inference process. The final posterior 
distribution is then, after following the standard
procedure in Bayesian statistics of including a prior for the newly defined random 
variables, given by
\begin{equation}
p(\mathbf{w},\alphabold,\sigma^2|\mathbf{d}) = \frac{p(\mathbf{d}|\mathbf{w},\sigma^2) 
p(\mathbf{w}|\alphabold)p(\alphabold)p(\sigma^2)}{p(\mathbf{d})},
\label{eq:bayes2}
\end{equation}
where we have used the fact that the likelihood does depend directly on $\mathbf{w}$ 
and not on the particular choice of $\alphabold$,
and that the priors for $\alphabold$ and $\sigma^2$ are independent.

The transformation to a hierarchical approach allowed \cite{Tipping} to
regularize the regression problem by favoring the sparsest solutions, i.e., the solution
that contains the least number of non-zero elements in $\mathbf{w}$. This is done by
defining the following prior
\begin{equation}
p(\mathbf{w}|\alphabold) = \prod_{i=1}^{4C+1} \mathcal{N}(w_i|0,\alpha_i^{-1}),
\end{equation}
where $\mathcal{N}(w|\mu,\sigma^2)$ is a Gaussian distribution on the variable $w$ 
with mean $\mu$
and variance $\sigma^2$. When using an appropriate hyperprior $p(\alphabold)$ (for instance, 
a sufficiently
broad Gamma distribution), the marginal prior $p(\mathbf{w})$ obtained by integrating 
out the $\alphabold$
parameters strongly favors very small values of $\mathbf{w}$, leading to a very sparse 
solution. The values of $\alphabold$ are estimated from the data using a Type-II maximum 
likelihood approach \cite[see][for details]{Tipping}.

Applying this scheme, we find the values of the few non-zero elements of the
vector $\mathbf{w}$, together with their confidence intervals obtained with a set of
calibration stars. Once these parameters
are fixed, the next step is to apply the inferred model to a set of test stars and compare the calculated 
temperatures with the tabulated ones.

\begin{figure*}[t!]
\centering
\includegraphics[width=\columnwidth]{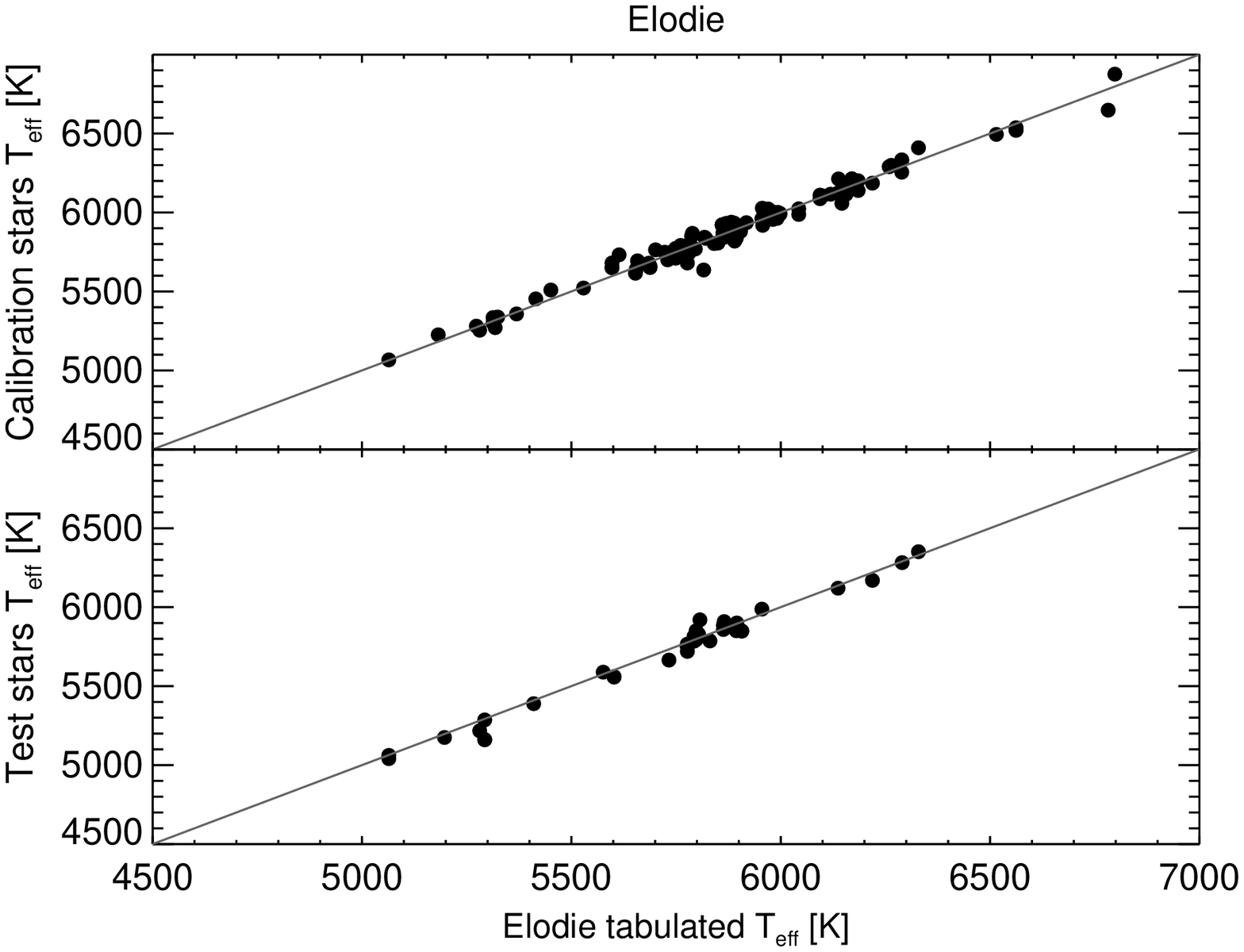}
\includegraphics[width=\columnwidth]{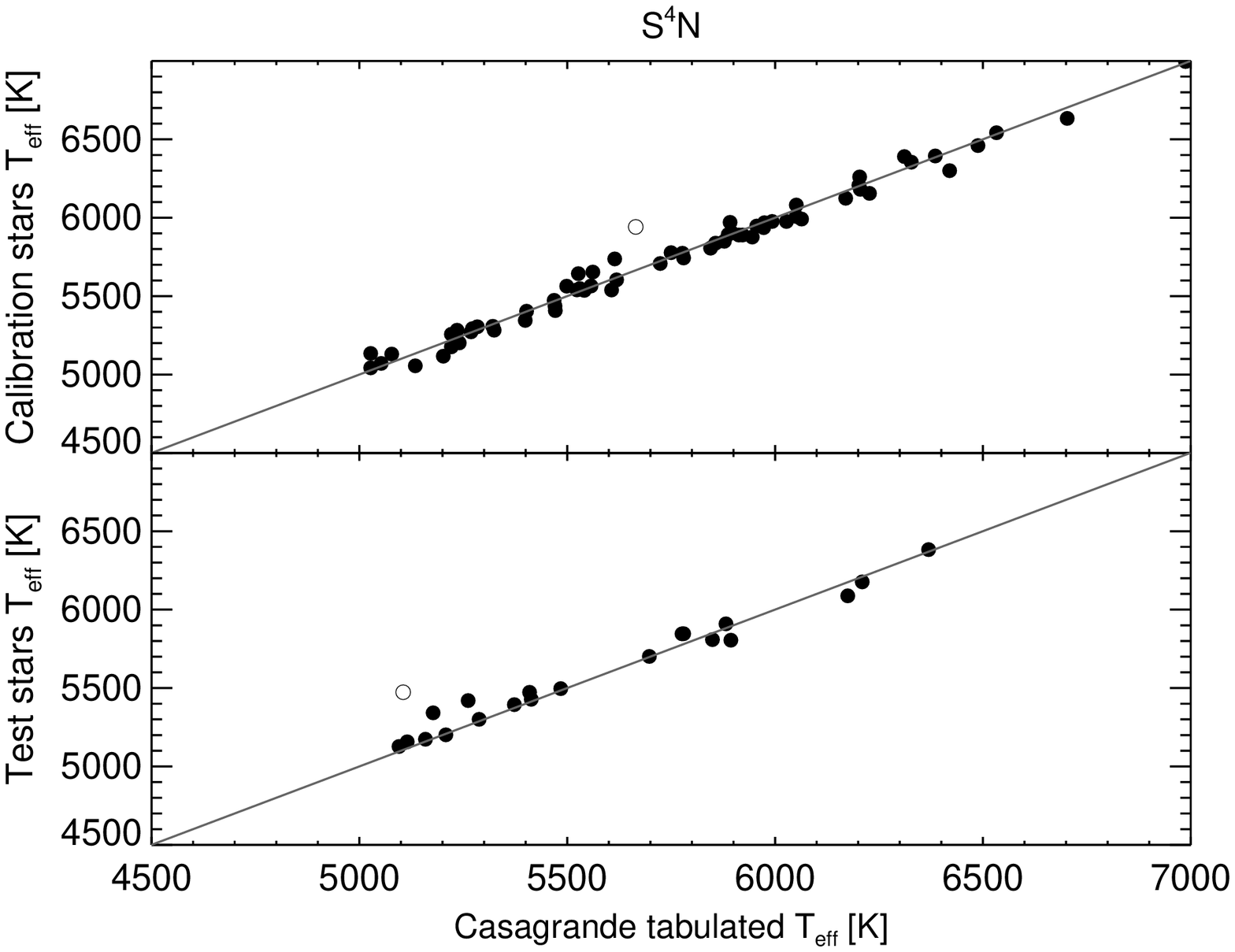}
\caption{Calculated temperatures vs tabulated temperatures for the calibration group of stars (upper panel) and the
test group of stars (lower panel) of the Elodie sample.}
\label{Elodie_S4N_tcalib}
\end{figure*}

\section{Results}
\label{Results}

We now discuss the results of applying the previous formalism to the
calibration of effective temperature. The coefficients $\mathbf{w}$
are computed using a set of reliable stars for which we have  good
estimates of the temperature  and apply the model to a set of stars 
of unknown temperatures. We first apply that calibration internally 
to Elodie and S$^4$N (calibration and test spectra from the same library), 
and then externally (calibration and test spectra from different libraries).

It is interesting to point out that, when performing the internal calibration for just one 
library, we found that the spectral range that we used was not critical for 
obtaining 
reliable results. Similar  results were obtained for different spectral 
ranges, unless they were too small. In any case, and in order to avoid further complications, 
we use the spectral range chosen in Sec. Sect. \ref{Connection}.

\subsection{Elodie}
\label{Elodie2}

As noted above, we considered 1245 spectra of 941 different stars that satisfied
all the selection  criteria described 
in Sect. \ref{The spectral libraries}. Of those stars, only 159 have been flagged as 
having the highest  quality ($Q_{T\mathrm{eff}}=4$), 
temperatures between 5000 K and 7000 K, and values of $\log g \geq 3$. 



\begin{table}[t!]
\caption{Internal calibration with stars from Elodie and S$^4$N}
\label{Elodie tab 1}
\begin{center}
\begin{tabular}{c c c c}
\hline\hline
\mbox{Description} & \mbox{N. PCs} & \mbox{calib. rms} (K) & \mbox{test rms} (K) \\
\hline
Elodie internal & 2 & 75  &  70 \\
& 3 & 64  &  67 \\
& 4 & 59  & 55  \\
& 5 & 49 & 55 \\
& 7 & 44 & 45 \\
\hline
Elodie internal & 3 & 55  &  111 \\
with $Q_{T\mathrm{eff}}=3$ & 4 & 49  & 107  \\
& 7 & 47 & 97 \\
& 9 & 46 & 98 \\
\hline
S$^4$N internal & 2 & 85 & 66 \\
& 4 & 81 & 62 \\
& 5 & 66 & 60\\
& 6 & 63 & 54\\
& 14 & 61 & 53  \\
\hline
\hline
\end{tabular}
\end{center}
\end{table}

Our first test is to calibrate with a subset of those spectra 
(the first 121, ordered by HD number), calculate the temperature of 
the rest of them (the remaining 38), and compare the calculated temperatures 
with the ones tabulated in the Elodie database. The results are summarized
in Table 1, where we show the rms residual between the predicted
temperatures and the original ones in the database as a function of the
number of principal components included in the calibration of Eq. \ref{eq:calibration}.
This Table presents the rms for the calibration with 159  well
as the results from experiments using only spectra with $Q_{T\mathrm{eff}}=3$ 
(since those are the second best spectra in quality) for different number of PCs.
One of the key results is that the rms for the test stars (those not used to build
the calibration) is very similar to the rms for the calibration stars when only 
the $Q_{T\mathrm{eff}}=4$ stars are considered. This indicates   
that the regression 
is not overfitting the calibration data. 
An additional proof of this is that an accurate calibration
is obtained independently from the number of PCs we use. The left-hand 
panel of Fig. \ref{Elodie_S4N_tcalib} shows the tabulated
$T_\mathrm{eff}$ versus the predicted temperature for the calibration stars (upper panel) and 
test stars (lower panel).


It is clear from Table 1 that the rms values for the test stars
with reduced quality ($Q_{T\mathrm{eff}}=3$) are higher than for the 
highest-quality stars, while this is not the case for the calibration stars. 
We interpret this as an indication that our calibration based only on stars
with the highest quality temperatures is reliable, while this is not the
case when including $Q_{T\mathrm{eff}}=3$ stars.  
Once we have verified that the method works, we use all the stars 
with $Q_{T\mathrm{eff}}=4$ as calibrators and infer the temperature of 
the sample (the complete set of 630 stars that passed all the filters and have 
tabulated temperatures between 5000 K and 7000 K). 
By doing so, we infer temperatures of $Q_{T\mathrm{eff}}=4$ quality for all of them. 
Table 2, includes the calculated temperatures for 
all the stars.

\subsection{S$^4$N}
\label{S4N2}

A total of 119 spectra of 119 different stars are available for this catalog. 
Once we apply the effective temperature and surface gravity filters 
we are left with 86 stars. Since we assume that all the $T_\mathrm{eff}$ are equally 
well determined, we use the first 65 stars as a calibration set and the remaining 
21 as a testing sample. As explained in Sect. \ref{S4N}, we adopt the $T_\mathrm{eff}$
values we calculate for these stars using the \citet{Casagrande} $b-y$ 
calibration.


The right panel in Fig. \ref{Elodie_S4N_tcalib} shows the tabulated reference 
temperatures vs. those we calculate from our method for both the calibration (upper panel)
and testing (lower panel) stars. As illustrated in the figure, there
are two stars (one in the calibration set and the other in the testing set) 
for which the predicted
temperatures differ sinificantly from the tabulated temperature (marked with an open circle).
Those two stars are HR 7578,  a spectroscopic binary 
\citep{Fekel}, and HD 188512 
(Alshain),  a variable star that is evolving off the main sequence \citep{Alshain}.


A summary of the results obtained is shown in Tab. \ref{Elodie tab 1}
(the rms have been calculated without taking into account the two stars mentioned above).
The calibration and test rms values are very similar again, which suggests that the model is 
not overfitting the calibration data 
and that it is valid for all the spectra in the sample. 
The slightly larger values of the calibration rms as compared with the test rms 
is probably due to an small overestimation of the noise in the input data. 
Nevertheless, this does not strongly affect the calibration.

\subsection{External application}
\label{External applications}

We have tested that our method works internally 
both for Elodie and for S$^4$N independently. 
Now we check whether we can use one library to calibrate 
the effective temperatures of the other one and homogenize the scales. 
We calibrate with Elodie (using just the $Q_{T\mathrm{eff}}=4$ spectra, 
in the range of 5000-7000 K and $\log g \ge 3$) and infer the 
temperatures of the stars in S$^4$N.

We have compared the stars that Elodie and S$^4$N 
have in common (out of 29 mentioned above there are 27 in 
the correct range of temperature and $\log g$). 
The mean difference between the 
Elodie tabulated temperatures and the ones calculated 
projecting S$^4$N onto Elodie is 88 K, with an rms of just 40 K. 
In other words, the inferred temperatures are highly correlated 
with the Elodie ones but with a systematic offset of 88 K.




\section{Temperature Scales}
\label{Temperature Scales}

We have seen that both Elodie and S$^4$N provide internally coherent 
temperature scales, and that our method is able to infer the 
temperature of a star, given its spectrum, with a reasonably small error. 
We have also seen that when we attempt to apply a PCA calibration 
based on Elodie spectra to S$^4$N spectra,  a systematic
offset between the derived temperatures appears. 
In this section we examine whether the temperature scales used to
calibrate our analysis of Elodie spectra, i.e. the Elodie $Q_{T\mathrm{eff}}=4$
literature-based temperatures, and those used for our analysis of the
S$^4$N spectra, i.e. the $B-V$/$b-y$ \citet{Casagrande} IRFM-based calibrations,
are compatible. We also compare with three additional sources of 
effective temperatures: 
the original scale of \citet{AlonsoDwarfs} adopted by \citet{Allende} in S$^4$N, 
the spectroscopic temperatures derived by \citet{Elodie1} from Elodie spectra, 
and the direct determinations by \citet{Cayrel}.

We selected 
the set of 29 stars that Elodie and S$^4$N have in common, 
and  calculated the mean difference and the rms between the  
effective temperatures from \cite{AlonsoDwarfs} or  \cite{Casagrande}, 
and those from Elodie. 
We find that, on average, the Elodie temperatures are higher than those
of \cite{AlonsoDwarfs} by 70 K (with an rms scatter of 87 K), while they are cooler
than those of \cite{Casagrande}  by 28 K (rms scatter of 85 K) when
their $B-V$ calibation is used, or by 50 K (rms scatter of 58 K) when
their $b-y$ calibration is adopted. We
embraced the \cite{Casagrande} $b-y$ calibration for the analysis of
S$^4$N spectra due to the smaller scatter found in this comparison.
Thus, the warmest scale is that by \cite{Casagrande}, followed by the
one based on classical high-resolution spectroscopic 
analyses, Elodie or \cite{Elodie1}, and finally the one proposed 
by \cite{AlonsoDwarfs}.



The  Elodie $Q_{T\mathrm{eff}}=4$ or spectroscopic scale (and hence our method, 
which uses it for calibration), seems to be in good agreement with 
the recent results of \cite{Cayrel}. These authors have calculated the effective temperature
for 11 stars derived from angular diameters and bolometric fluxes, a technique which is expected
to give the most fundamental measurement of the effective temperature. 
All their 11 stars are in S$^4$N, and 7 are also in Elodie. 
Table 3 displays the \cite{Cayrel} direct effective temperatures, 
the \cite{Elodie1} and \cite{Casagrande} temperatures, those adopted in the original
S$^4$N paper obtained applying the \cite{AlonsoDwarfs} calibrations, as well as the 
temperatures calculated using the internal PCA calibration of Elodie. 
The \cite{Casagrande}  $b-y$ temperatures are,  on average, 59 K 
warmer than the \cite{Cayrel} direct temperatures, whereas 
both \cite{Casagrande} $B-V$ temperatures and the 
the \cite{Elodie1}
temperatures are consistent with the \cite{Cayrel} temperatures, 
and the \cite{AlonsoDwarfs} values are somewhat lower.

\begin{figure}[t!]
\centering
\includegraphics[width=\columnwidth]{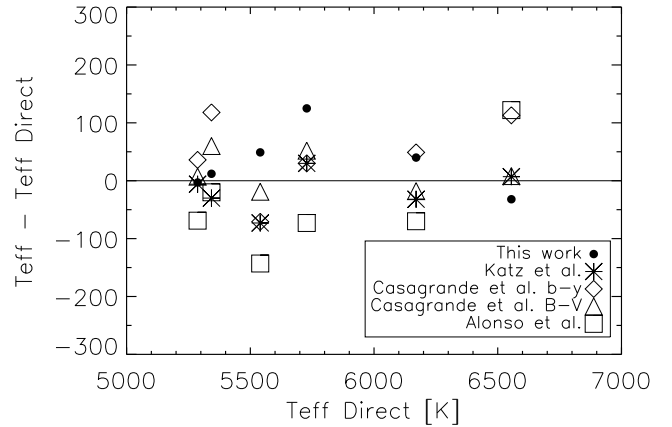}
\caption{Differences between different estimates of the effective temperature
and the direct values for six stars in common for the sources compared in
Table \ref{Cayrel test}.}
\label{sixstars}
\end{figure}

Excluding the Sun, which is usually assigned a temperature of 5777 K, 
there are 6 stars in common for all four 
sources compared in Fig. \ref{sixstars} and Table 3. The mean (and standard
deviation) of the differences between each of the following sources
and the direct temperatures of \cite{Cayrel} for these stars are:  
$-17 (36)$ K for \cite{Elodie1}, $+46 (69)$ K for \cite{Casagrande} $(b-y)$, 
$+15 (34)$ K for \cite{Casagrande} $B-V$, 
$-42 (90)$ K for \cite{AlonsoDwarfs}, and $+32 (54)$ K for
our PCA calibration of Elodie spectra, consistent with the discussed systematics.



\addtocounter{table}{1} 

\begin{table*}[b!]
\caption{Comparison of $T_\mathrm{eff}$ in K in different catalogs. The solar values are 
forced to 5777 K for some sources.}
\label{Cayrel test}
\begin{center}
\begin{tabular}{l c c c c c c}
\hline\hline
\mbox{Star} & \mbox{Direct} & \mbox{Elodie} & \mbox{S$^4$N} & \mbox{S$^4$N} & \mbox{S$^4$N} & \mbox{Elodie} \\ 
\mbox{} & \mbox{} & \mbox{Katz et al.} & \mbox{Casagrande et al. $b-y$} & \mbox{Casagrande et al. $B-V$} & \mbox{Alonso et al.} & \mbox{Our calibration} \\
\hline
Sun        & 5777 $\pm$ 5         & 5777 & 5777 & 5777 & 5777 & 5750 \\ 
$\mu$ Cas  & 5343 $\pm$ 18        & 5313 & 5461 & 5403 & 5323 & 5355  \\
$\upsilon$ And  & 6170 $\pm$ 18   & 6138 & 6219 & 6152 & 6100 & 6210 \\
$\tau$ Cet & 5376 $\pm$ 22        & -    & 5451 & 5377 & 5328 & - \\
$\epsilon$ Eri & 5107 $\pm$ 21    & -    & 5159 & 5085 & 5052 & -\\
$\alpha$ CMi A & 6555 $\pm$ 17    & 6562 & 6668 & 6563 & 6677 & 6523\\
$\beta$ Vir & 6062 $\pm$ 20       & -    & 6176 & 6112 & 6076 & -\\
$\eta$ Boo & 6019 $\pm$ 18        & -    & 6088 & 6053 & 5942 & -\\
$\zeta$ Her & 5728 $\pm$ 24       & 5758 & 5758 & 5780 & 5655 & 5853 \\
$\mu$ Her & 5540 $\pm$ 27         & 5467 & 5469 & 5521 & 5397 & 5589  \\
$\sigma$ Dra & 5287 $\pm$ 21      & 5281 & 5323 & 5294 & 5218 & 5284 \\ 
\hline
\end{tabular}
\end{center}
\end{table*}

\section{Application to the Elodie archive}
\label{allelodie}

We have mainly based our analysis on the Elodie library spectra made public 
by \citet{ElodieB}, and \citet {ElodieA}, but in the 12 years (1994-2006) that this 
instrument was in operation, many more data were gathered, and these
are publicly available from the Elodie archive \citep{ElodieC}.

We have downloaded 34,033 spectra from the Elodie archive, selected
those in our range of interest (F- and G-type stars), 
and applied our PCA calibration to infer 
effective temperatures for them. We keep only 18,696 spectra for which the 
PCA components are within the range of our calibration, and for which the 
PCA reconstruction matches the original to better than 5\%. The 
derived $T_{\rm eff}$ values are provided in Table 4, available only in 
electronic form.

The 18,696 spectra we provide effective temperatures for correspond to 
4,039 unique stars, of which there are 2,553 with just one spectrum.
The remainder 1,486 stars have between 2 and 368 spectra, with a median 
of 4 spectra per star. The derived effective temperatures for the 1,486 
stars with multiple spectra show  a mean rms scatter 
of 32 K with a standard 
deviation of 58 K, and a median rms scatter of 18 K. 
Clearly the temperatures we provide 
for different Elodie spectra
of the same star are in general highly consistent.
 
Table 5, only in electronic form, provides a single temperature for each
object in the Archive, averaging the results when more than one spectra are
available. Note that the archive includes some non-stellar spectra.

 Our method is well-suited for application to very large sets 
of stellar spectra obtained as part of projects such
as SEGUE/SDSS (Yanny et al. 2009), APOGEE (Eisenstein et al. 2011), 
RAVE (Siebert et al. 2011), GALAH (Zucker et al. 2013), Gaia-ESO (Gilmore et al. 2012), 
or Gaia (Lindegren et al. 2008). This only requires 
observations with the survey instrumentation of a suitable calibration
sample, which may be a difficult enterprise for programs focusing on faint stars,
since the best reference stars are quite bright.

\section{Conclusions}
\label{Conclusions}

We have developed a method to obtain effective temperatures 
from low-resolution spectroscopic data. We 
project the observed spectra onto the eigenvectors 
 and use a calibration curve derived using a
robust non-parametric regression on a set of stars with reliable temperatures.

Unlike the photometric or spectrophotometric methods, 
this procedure does not suffer from systematic errors associated with 
interstellar reddening. Also, given a certain $T_\mathrm{eff}$
scale the method provides coherent $T_\mathrm{eff}$ values on a homogeneous for all the sample.
For instance, within the Elodie library, we carried out the calibration with only 159 spectra with well determined
temperatures and obtained the temperatures of other 630 spectra with the same
quality (see  Table 2). 
We applied as well the calibration to nearly 19,000 spectra of some 4,000 unique
stars from the Elodie archive.

We checked the method internally for both Elodie and S$^4$N spectra 
with excellent results. 
However, when applying the Elodie calibration to S$^4$N spectra, we 
discovered that the method was overestimating the S$^4$N temperatures by about 90 K.
We also find that the IRFM-based \cite{Casagrande} 
$B-V$ scale is very close to the Elodie (spectroscopic) scale,
 but that is not the case for their $b-y$ calibration.
We compared those scales with the direct temperatures provided 
by \cite{Cayrel} and found that the Elodie scale and the \cite{Casagrande} 
$B-V$ calibration are in good 
agreement with them, whereas the \cite{Casagrande} $b-y$ calibration presents an 
offset of about $+50$ K.

One could use a third spectral library to check 
whether the method works properly in the spectral range used. 
We tried introducing the solar spectrum from  \citep{Kurucz},
  smoothed appropriately, 
and the temperature we obtained was 5789 K, which is in very 
good agreement with the real temperature, 
while the method returned a higher temperature for the S$^4$N 
solar spectra 
(in line with the  $\sim 80$ K offset expected from tests with
stars in common between Elodie and S$^4$N).

It would be useful to expand the set of 
reference (direct) $T_\mathrm{eff}$ values. 
We have tried using the temperatures provided by \cite{Cayrel} 
but there was not enough  information for a successful PCA mapping; the 
minimum number of calibration spectra required is about 
50 spectra for our $T_\mathrm{eff}$ range, or 
approximately one star per 40-K interval.

In addition to the practical application of our Elodie-based calibration 
of effective  temperature, the most interesting outcome of this study 
is that our experiments demonstrate the potential of PCA to extract
information from stellar spectra, and in particular a close connection 
between the most important principal components and the stellar
effective temperatures.

\begin{acknowledgements}
We are thankful to Luca Casagrande for useful comments on the manuscript. AAR acknowledges financial support by the Spanish Ministry of Economy and 
Competitiveness through projects AYA2010-18029 (Solar Magnetism and Astrophysical 
Spectropolarimetry) and Consolider-Ingenio 2010 CSD2009-00038. AAR also 
acknowledges financial support through the Ram\'on y Cajal fellowship.
JMB acknowledges financial support provided by the IAC summer research grants.
\end{acknowledgements}


%
%
%

\bibliographystyle{aa}
\bibliography{bibliography}

\onllongtab{2}{
\begin{longtable}{lcccccccl}
\caption{Effective temperatures for stars in the Elodie library.}\\
\hline
\hline
Filename & Star & $T_{\rm eff}$ & $Q_{T\mathrm{eff}}$ & $\log g$ & $Q_{\log g}$ & [Fe/H] &
$Q_{\rm [Fe/H]}$ &  $T_{\rm eff}$ This work \\
\hline
\endfirsthead
\caption{Continued.} \\
\hline
Filename & Star & $T_{\rm eff}$ & $Q_{T\mathrm{eff}}$ & $\log g$ & $Q_{\log g}$ & [Fe/H] &
$Q_{\rm [Fe/H]}$ &  $T_{\rm eff}$ This work \\
\hline
\endhead
\hline
\endfoot
00001.fits & HD  245 & 5433. & 3 & 3.50 & 1 & -0.76 & 3 & 5739. \\
00003.fits & HD  400 & 6146. & 4 & 4.09 & 1 & -0.28 & 4 & 6073. \\
00004.fits & HD  693 & 6156. & 4 & 4.13 & 1 & -0.42 & 3 & 6123. \\
00010.fits & HD  3268 & 6130. & 3 & 4.01 & 1 & -0.24 & 2 & 6146. \\
00012.fits & HD  3567 & 5991. & 4 & 3.96 & 1 & -1.25 & 4 & 5977. \\
00013.fits & HD  3628 & 5701. & 4 & 4.06 & 1 & -0.19 & 2 & 5768. \\
00020.fits & HD  4614 & 5890. & 4 & 4.40 & 1 & -0.28 & 4 & 5909. \\
00028.fits & HD  6582 & 5313. & 4 & 4.40 & 1 & -0.83 & 4 & 5356. \\
\hline
\end{longtable}
}

\end{document}